\documentclass[%
 reprint,
 amsmath,amssymb,
 aps,
prb,
]{revtex4-2}

\usepackage{graphicx}
\usepackage{dcolumn}
\usepackage{bm}
\usepackage{xcolor}
\usepackage{ulem} 

\newcommand{\updownarrows}{\mathbin\uparrow\hspace{-.0em}\downarrow}

\begin{document}


\title{Optical Controllable Spin-Polarization in Two Dimensional Altermagnets via Robust Spin-Momentum Locking Excitons}
\author{Jiuyu Sun}
\author{Jinzhe Han}
\author{Yongping Du}
\email{njustdyp@njust.edu.cn}
\author{Erjun Kan}
\email{ekan@njust.edu.cn}
\affiliation{MIIT Key Laboratory of Semiconductor Microstructure and Quantum Sensing, Nanjing University of Science and Technology, Nanjing 210094, China}

\date{\today}

\begin{abstract}

Spin-momentum locking (SML) excitons in two-dimensional semiconductors are appealing to programmable optical control of spin-polarized carriers in ultrafast spintronics.
To address the current thirsty for long-lived excitons with zero-external-field stability and room-temperature spin-polarization, we hereby predict the existence of intrinsically SML excitons in altermagnetic V$_2 X_2$O ($X=$ S, Se) driven by giant non-relativistic spin-splittings ($>$ 1.2 eV).
First-principles calculations reveal SML excitons with binding energies exceeding 1400 meV in monolayers and 430 meV in their van der Waals heterobilayers, along with stacking-dependent optical selection rules for tunable interlayer excitons.
These remarkable physical properties, combined with their long radiative lifetimes, strongly suggest the feasibility of SML excitons with robust spin-polarization at room temperature. Our work provides a new paradigm for SML exciton physics via the novel altermagnetism, opening up new possibilities for all-optical manipulation in advanced opto-spintronics.

\end{abstract}

\maketitle


\section{introduction}
The optical control of spin-polarized carriers in two-dimensional (2D) semiconductors has emerged as a pivotal area of research, offering ultrafast manipulation of spin and charge and non-contact operation through polarized light\cite{Schaibley2016review,2D-Polarized-Materials2021,2023review2DwithLight,ZHANG2025review}.
Recently, spin-momentum locking (SML) excitons in 2D semiconductors attract considerable attention, which are bound electron-hole pairs with spin orientation intrinsically tied to crystal momentum\cite{HongiyiYu2015review,Rivera2016,Mak2018,Ciarrocchi2022,Xu2024valleyReview}.
SML excitons, in principle, bypass the need for external magnetic fields by leveraging momentum-dependent spin textures, offering deterministic optical addressability, extended spin lifetimes, and robust quantum coherence, which are critical for advancing energy-efficient spin logic and room-temperature quantum technologies.

Although remarkable progresses have been made via an intensive pursuit of SML excitons, significant challenges remain in their practical implementation.
In conventional non-magnetic 2D materials like transition metal dichalcogenides (TMDCs), e.g. MoS$_2$, the generation of SML excitons via circularly polarized light replies on the spin-orbit coupling (SOC) induced spin-splitting\cite{Cao2012nc-TMDvalley,Wang2018RMPexcitons,Rivera2018-IX,Jiang2021-IXreview,Xu2024valleyReview,Yambo-SpinorExciton}.
However, the spin-splitting is inherently weak, on the order of tens of meV, unless toxic heavy elements are used\cite{Selig2019,Helmrich2021,Chen2022,Shudong2024-prb}. The consequent difficulty for spin-selective identification and manipulation, as well as the rapidly decay of spin-polarization at room temperature, require cryogenic temperatures or external stimuli for stabilization\cite{Yu2014,Glazov2015reviewExciton,Thygesen2016,Wang2018valleyRealx,Guo2019,Xu2021,JiangXiang2021}.
These challenges expose the fundamental limits of relativistic mechanisms in achieving robust spin polarization in non-magnetic 2D semiconductors.

Efforts to overcome these issues have turned to the recently discovered 2D magnetic semiconductors\cite{Jiang2019,Heissenbuttel2021,Dirnberger2022,DiSabatino2023,NiBr2Exciton2024}, which host long-lived spin-polarized excitons due to their intrinsic long-range magnetic orders. 
For instance, spin-polarized excitons are reported in chromium trichlorides (ferromagnetic) and transition metal phosphorous trichalcogenides (antiferromagnetic), with large binding energies in theory\cite{Wu2019,LiYangMag2020,Birowska2021,Molina-Sanchez2020,Wu2022} and clear spin-polarized optical responses in experiment\cite{Seyler2018,Hwangbo2021}.
Yet, challenges persist: the ferromagnetism only supports spin-polarized states in the spin-majority direction and can hardly survive in 2D semiconductors at room temperature; the lack intrinsic spin-splitting in antiferromagnets makes it difficult to achieve efficient optical control of spin-polarized excitons without additional symmetry-breaking operations.
These trade-offs underscore a critical demand for a material platform that intrinsically combines considerable non-relativistic spin-splitting, room-temperature stability, and efficient optical control.

\begin{figure}[htb]
\includegraphics[width=7.5 cm]{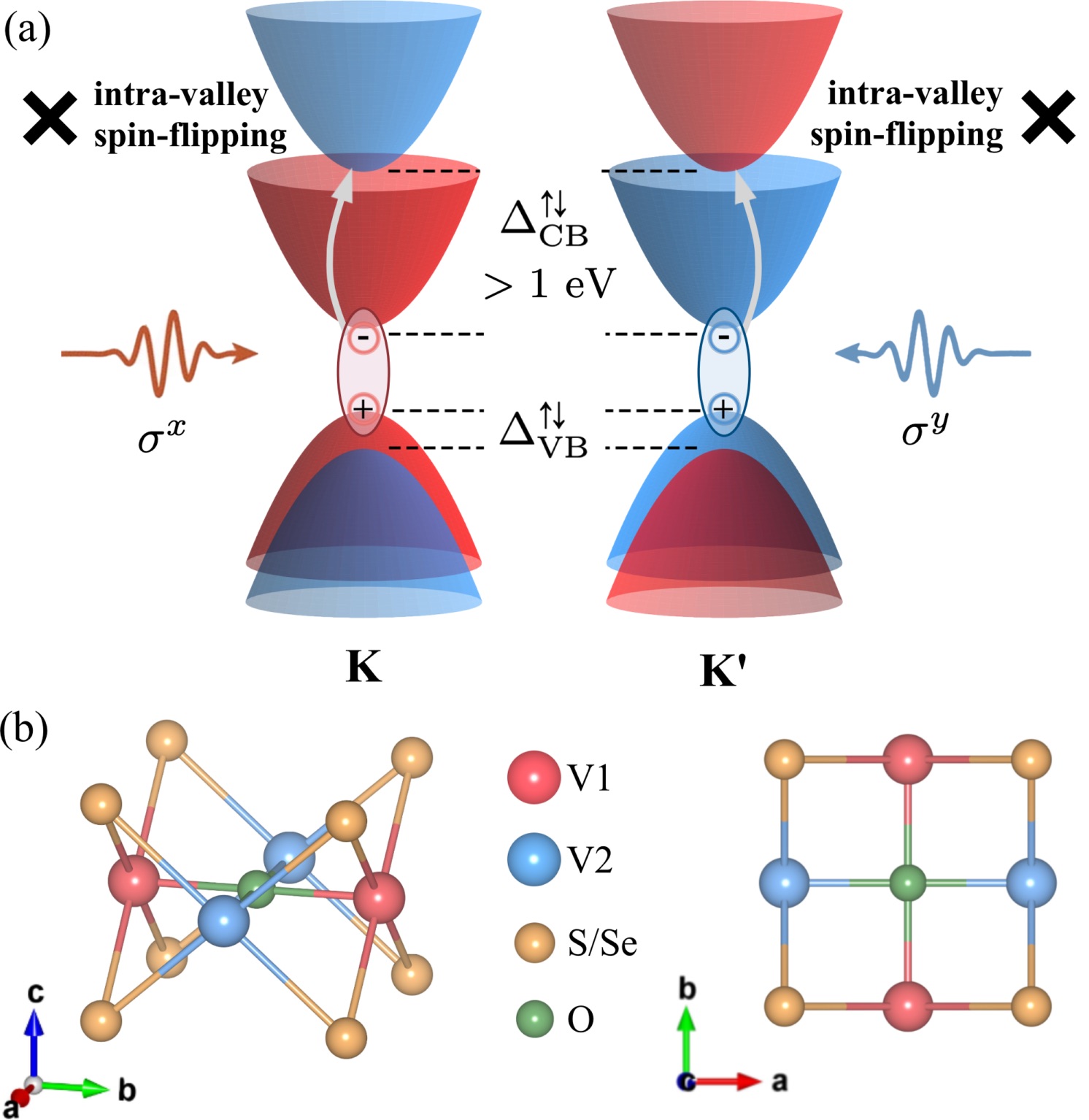}
\caption{\label{fig:1} Schematic diagram of the generation of SML excitons at $K$ ($X$) and $K'$ ($Y$) valleys linear polarized light along $x/y$ direction ($\sigma^x/\sigma^y$), where the large exchange-induced spin-splittings for both valence and conduction bands ($\Delta^{\updownarrows}_{\mathrm{VB}}$ and  $\Delta^{\updownarrows}_{\mathrm{CB}}$, over 1 eV in monolayer V$_2 X_2$O eV) would remarkably reduce the intra-valley spin-flipping, e.g. for conduction bands. (b) Atomic structure of monolayer V$_2 X_2$O with V1 (spin-up V), V2 (spin-down V), S/Se, O atoms colored in red, blue, orange and green, respectively.}
\end{figure}

The novel discovered altermagnets uniquely combine thermal stability as antiferromagnets--zero net magnetization and ferromagnetic-like non-relativistic spin-splittings\cite{Smejkal2020,Ma2021,Smejkal2022a,Zhu2024,Krempasky2024,Song2025AltermagReview,ZhouTong2025PRL,ZhouTong-Zhu2025NL}. 
Their symmetry-protection induced momentum-dependent spin texture\cite{Yu2020,Ma2021,Wu2024,Zhang2024,Qi2024}
enables a momentum-dependent and spin-selective optical response\cite{weber2024opticalexcitationspinpolarization,weber2024ultrafastdynamicsaltermagnetic,Eskandari2025prb-model-td-AM,zhouzhaobo2025ultrafastAM,JMoore2025prb-AMmodel}. 
While these previous works\cite{weber2024opticalexcitationspinpolarization,weber2024ultrafastdynamicsaltermagnetic,Eskandari2025prb-model-td-AM,zhouzhaobo2025ultrafastAM,JMoore2025prb-AMmodel} focus on the single-particle optical properties in mainly in bulk altermagnets, we theoretically propose that SML excitons can be generated in 2D altermagnetic semiconductors\cite{Ma2021,Qi2024,Wang2025,Zeng2024a,Wu2024}, present a promising material platform towards room-temperature robust spin-polarized excitons.
Exemplified by the successfully synthesized and verified \cite{Lin2018,Zhang2025,Jiang2025} V$_2 X_2$O ($X=$ S, Se, Te) materials, our first-principles calculations reveal that 2D V$_2 X_2$O ($X=$ S, Se) systems host intrinsically SML excitons with binding energies exceeding 1400 meV in monolayers and 430 meV in their van der Waals heterobilayer (vdWH). 
In addition, we show that perfect spin-polarization of the SML excitons upon linear polarized light is enabled by the momentum selective linear dichroism\cite{JMoore2025prb-AMmodel} in monolayers, as well as stacking-dependent optical selection rules for tunable interlayer SML excitons in the vdWH.
More importantly, the giant non-relativistic spin-splittings, exceeding 1.2 eV in V$_2 X_2$O, could drastically reduce the spin-valley depolarization of the SML excitons via various routines, e.g. intra-valley spin-flipping in Figure~\ref{fig:1}(a).
This altermagnetic mechanism transcends the limitations of SOC-driven systems and conventional magnets, enabling spin logic operations without cryogenic or external-field constraints.

\section{Theoretical Methods}

\subsection{First-principles calculations}

Our density functional theory calculations were performed using the Quantum Espresso package\cite{QE} with the optimized norm-conserving Vanderbilt pseudopotentials\cite{ONCV-psp,PseudoDojo}.
All geometric structures were fully relaxed with the PBE\cite{PBE} functional, and the lattice constant of unit cell of V$_2$Se$_2$O monolayer was optimized to be 3.88~\AA, which is in agreement with previous works\cite{Ma2021,Lin2018,Qi2024}. The Néel temperatures of the monolayers are estimated\cite{OPENMX0,*OPENMX1,*OPENMX2,*OPENMX3,*TB2J,*Jx,*mcsolver} to be over 700 K.
For the quasiparticle band structures, we firstly calculated the ground state with PBE+$U$ with different values of $U_{\rm eff}$, followed by single-shot $GW$ calculations with different recipes.
After a comprehensive evaluation of various methods\cite{pbe0,hse06,Jiang2010GW+U} (see Sec. III in the Supplemental Materials (SM)\cite{supp}), including PBE+$U$, $GW$@PBE+$U$, hybrid HSE06 and PBE0 functionals, we adopted the ground state by PBE+$U$ ($U_{\rm eff}=4.5$ eV) along with a scissor operator ($\Delta_\mathrm{sc}=2.0$ eV), as the quasiparticle bands in the next section and the input wavefunction for the following investigations on excitons and optical properties.
Note the main purpose of this work is to propose a new platform for studying SML excitons. In this sense, the employing of the scissor operator, to account for the quasiparticle self-energy effect on the electronic bands, intents to offer a reasonable estimation of the excitonic energies and optical spectra. In addition, one could refer to the results in SM\cite{supp} to learn the range of excitonic energies calculated with various methods.

Then the exciton properties (energies and wavefunctions) are obtained by solving the
Bethe-Salpeter equation (BSE) with a static dielectric screening\cite{BSE}:
\begin{equation}
\left(E_{c \mathbf{k}}^{\mathrm{QP}}-E_{v \mathbf{k}}^{\mathrm{QP}}\right) A_{v c \mathbf{k}}^{S}+\sum_{v^{\prime} c^{\prime} \mathbf{k}^{\prime}}\left\langle v c \mathbf{k}\left|K^{\mathrm{eh}}\right| v^{\prime} c^{\prime} \mathbf{k}^{\prime}\right\rangle A_{v^{\prime} c^{\prime} \mathbf{k}^{\prime}}^{S}=E_{\rm ex}^{S} A_{v c \mathbf{k}}^{S},
\end{equation}
where $A_{v c \mathbf{k}}^{S}$ is the $S$-th exciton wavefunction coefficient and $E_{\rm ex}^{S}$ is the $S$-th excitonic (excitation) energy. $K^{\mathrm{eh}}$ represents the many-body effect (electron-hole interaction) kernel. With the obtained excitons, the optical response Im[$\varepsilon(\omega)$] (imaginary part of the dielectric
function ) is computed as
\begin{equation}
{\rm Im}[\varepsilon(\omega)]=\frac{16\pi^{2}e^{2}}{\omega^{2}}\sum_{S}|\mathbf{e}\cdot\langle 0|\mathbf{v}|S\rangle|^{2}\delta\left(\omega-E_{\rm ex}^{S}\right),
\end{equation}
where $\mu_S=\langle 0|\mathbf{v}|S\rangle=\sum_{cv\mathbf{k}}A_{v c \mathbf{k}}^{S} \langle u_{v\mathbf{k}} \left| \mathbf{v} \right| u_{c\mathbf{k}} \rangle$ is the transition matrix element from the ground $|0\rangle$ to $S$-th excitonic state $|S\rangle$.
All the spin-resolved many-body perturbation calculations\cite{spinGW,GPP-HL,GPP-GN,haydock} were
performed by using BerkeleyGW package\cite{TruncationBGW,BGW}, and more computational details and convergence tests could be found in the SM\cite{supp}. 

Furthermore, the radiative lifetime of $S$-th exciton at temperature $T$ (K) is estimated via approach by Bernardi et al.\cite{Palummo2015a,RLifetime-Bernardi2019}:
\begin{equation}\label{tau_K}
    \langle\tau^S\rangle (T)=\tau^S_0 \frac{3}{4}{\left\{\frac{(E^S_{\mathrm{ex}})^2}{2M^S c^2}\right\}^{-1}}k_\mathrm{B}T ,
\end{equation}
with the radiative lifetime at zero-temperature:
\begin{equation}\label{tau_0}
    \tau^S_0 = \frac{A_{\mathrm{uc}}\hbar^2c}{8\pi e^2 E^S_{\mathrm{ex}}\mu_S^2},
\end{equation}
where $A_{\mathrm{uc}}$ is the area of the unit cell and $E^S_{\mathrm{ex}}$ is the excitation energy of the $S$-th zero momentum exciton with effective mass $M^S$. $\mu_S^2$ is the square of the dipole matrix element of the $S$-th exciton obtained by solving BSE.

\subsection{Generalized Wannier-Mott Model}

Based on the first-principles calculations, excitonic properties, such as binding energy and wavefunction could be also solved via the Wannier-Mott exciton model\cite{Wannier1937-exciton,Elliott1957}:
\begin{equation}\label{MW-model}
    \left[-\frac{\hbar^2}{2M^*}\nabla^2  + V(r) \right] \psi({\bf r}) = E^{\mathrm{W-M}}_{\mathrm{b}}\psi({\bf r}),
\end{equation}
where $M^*$ the effective mass of an exciton, i.e., $M^* = m^*_e m^*_h/(m^*_e+m^*_h)$, with $m^*_e$ and $m^*_h$ being the effective mass of excited electron and hole at target $K$-point, respectively. The eigenvalues $E^{\mathrm{W-M}}_{\mathrm{b}}$ are the energy levels of excitonic states, which are opposite to the excitonic binding energies. In a three-dimensional semiconductor with the dielectric constant $\varepsilon$, $V(r)$ is the screened Coulumb potential $-1/\varepsilon r$. However, due the nonlocal screening of electron-hole Coulomb interactions, such a hydrogenic model is hardly valid for the excitons in 2D finite-gap systems\cite{ShanJie2014prl-exciton,TonyHeinz2014prl-NonhydrogenicExciton,TonyHeinz2015NL-TMDExcitonExpt}. In this sense, we applied the well-known Rytova-Keldysh (RK) potential\cite{Keldysh1979}:
\begin{equation}\label{rk-pot}
   V_\mathrm{RK}(r) = -\frac{2\pi}{r_0}\left[   H_0(\frac{\kappa r}{r_0}) - Y_0(\frac{\kappa r}{r_0})\right],
\end{equation}
where $H_0$ is the Struve function, $Y_0$ is the Neumann function (also called the Bessel function of the second kind), and $\kappa$ is the dielectric environment $\kappa=1$ for free-standing layer(s). The screening length $r_0$ links with the 2D polarizability via $r_0=2\pi \alpha$\cite{Angel2011-2Dscreening}, and the latter can be obtained non-empirically via the Random Phase Approximation (RPA)\cite{Thygesen2016prl,Hieu2022-2DExcitonModel}:
\begin{equation}
  \alpha = \sum_{m, n} \int_{BZ} \frac{d\mathbf{k}}{(2\pi)^{2}} (f_{n\mathbf{k}} - f_{m\mathbf{k}}) \frac{\left| \left\langle u_{m\mathbf{k}} \left| \mathbf{r} \right| u_{n\mathbf{k}} \right\rangle \right|^{2}}{\varepsilon_{n\mathbf{k}} - \varepsilon_{m\mathbf{k}}},
\end{equation}
where the $f_{n\mathbf{k}}$, $\varepsilon_{n\mathbf{k}}$ and $u_{n\mathbf{k}}$ are the occupation number, electronic energy and wavefunction for $n$-th state at $\mathbf{k}$, respectively, which have been obtained from the PBE+$U$ calculations above.

\section{results and discussions}

\subsection{V$_2$S$_2$O monolayer}

\begin{figure}[htb]
\includegraphics[width=8.6 cm]{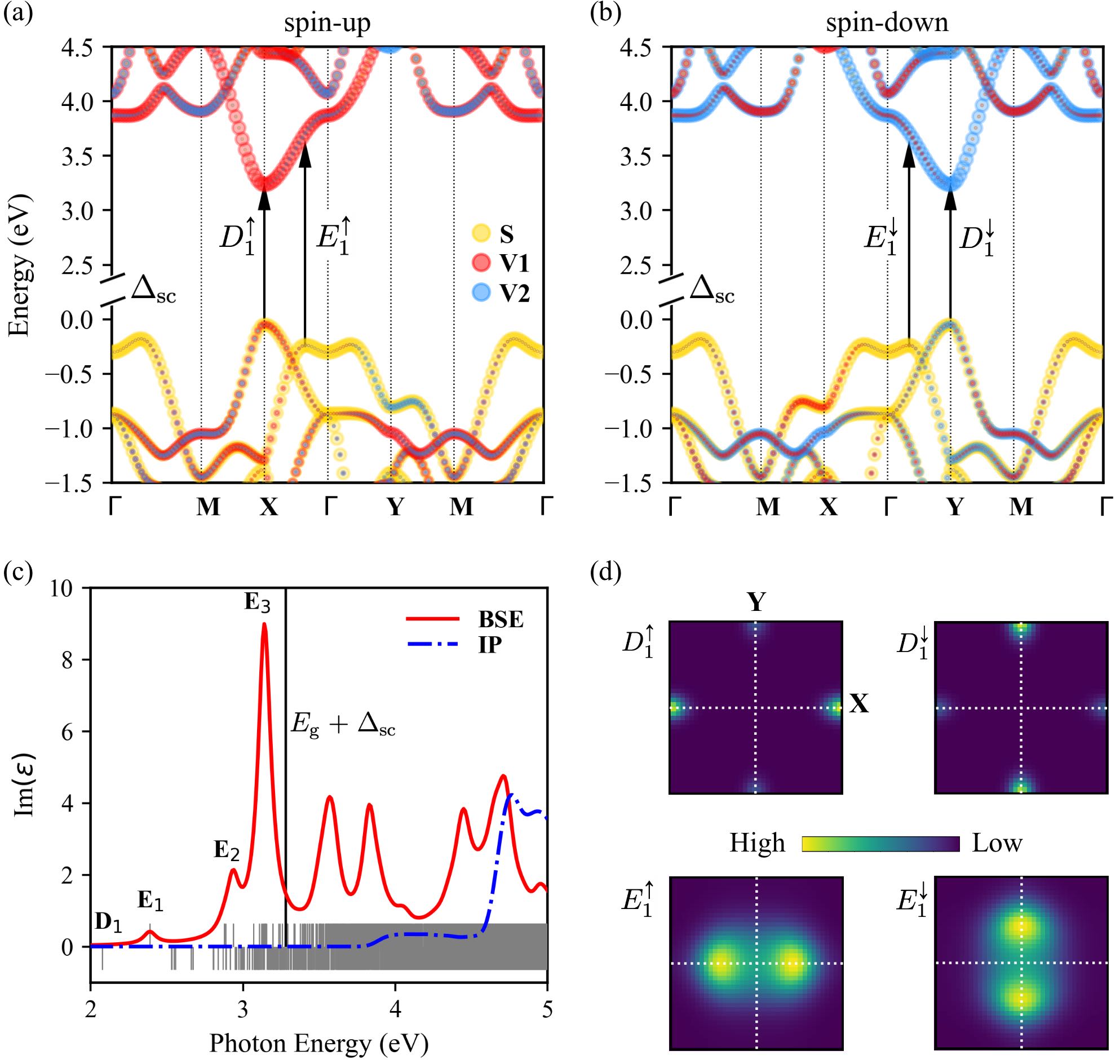}
\caption{\label{fig:2} Calculated results for the V$_2$S$_2$O monolayer. The (a) spin-up and  (b) spin-down band structures by PBE+U with a scissors operator ($\Delta_\mathrm{sc}=2.0$ eV, indicated by the y-axis break), projected onto the V1, V2 and S atoms with the weights proportional to the radii of the colored circles. Energy zero is at the top of valence band. The vertical arrows indicate the transitions of corresponding excitons. (c) Optical spectra Im($\varepsilon$) by BSE and IP for the light polarization along [110]. The vertical grey ticks indicate the excitonic energies with module square of dipole moment larger and smaller than $1\times10^{-2}$ are shown separately upon and below zero axis, respectively. The single-particle band gap after the rigid scissors operator ($E_\mathrm{g}+\Delta_\mathrm{sc}$) is indicated by the black vertical line.
(d) Reciprocal-space distributions of optical transitions for selected SML excitons.
}
\end{figure}

As shown in Figure~\ref{fig:1}(b), the V$_2 X_2$O monolayer has a square geometry with the collinear magnetic moments on the two V atoms aligning oppositely along the out-of-plane $z$-direction. We use V$_2$S$_2$O to illustrate the common electronic and optical properties for the monolayers, and more results for V$_2$Se$_2$O are available in the SM\cite{supp}. 
Calculated with PBE+$U_{\rm eff} = 4.5$ eV, the projected band structures (with a scissors operator $\Delta_\mathrm{sc}=2.0$ eV) for both spin channel are shown in Figure~\ref{fig:2}(a-b) and Table~\ref{tab:energy}. Due to the spin group $|C_2||C_{4z}|$, the bands are spin degenerate along $\Gamma$-$M$, whereas they are identical in energy but with opposite spin splittings along $\Gamma$-$X$ and $\Gamma$-$Y$.
There are two pairs of valleys around both the valence band maximum (VBM) and conduction band maximum (CBM) at $X$ and $Y$ points for spin-up and spin-down channels, respectively.  
A direct band gap of 1.28 eV at $X$ or $Y$ valleys is obtained with PBE+$U_{\rm eff} = 4.5$ eV.
Our single-shot $GW$ calculations, as well as the hybrid HSE06 and PBE0 functionals, show that the band gap and the spin splitting for VBM ($\Delta^{\updownarrows}_{\mathrm{VB}}$) largely depend on the computational methodology\cite{supp}.
On the contrary, the spin splitting at CBM valleys ($\Delta^{\updownarrows}_{\mathrm{CB}}$) is not much affected by the computational methods and retains a large $\Delta^{\updownarrows}_{\mathrm{CB}}$ over 1.1 eV, which is crucial for achieving robust spin-polarized excitons.

\begin{table}[htb] 
\caption{\label{tab:energy} Direct band gaps ($E_\mathrm{g}$) by PBE+$U$ with a scissor operator ($\Delta_\mathrm{sc}$) of 2.0 eV, spin-splittings in valence and conduction bands ($\Delta^{\updownarrows}_{\mathrm{VB}}$ and  $\Delta^{\updownarrows}_{\mathrm{CB}}$), the excitation energy of the lowest bright exciton $E_1$ ($E^1_{\mathrm{ex}}$) and the corresponding excitonic binding energy ($E^1_{\mathrm{b}}$) in monolayers and vdWHs in the unit of eV.} 
\begin{ruledtabular}
\begin{tabular}{rccccc}
  &  $E_\mathrm{g}+\Delta_\mathrm{sc}$ & $\Delta^{\updownarrows}_{\mathrm{VB}}$ &  $\Delta^{\updownarrows}_{\mathrm{CB}}$ & $E^1_{\mathrm{ex}}$  &  $E^1_{\mathrm{b}}$\\
\colrule
V$_2$S$_2$O & 3.28 & 0.76 & 1.27 & 2.39 &  1.55 \\
V$_2$Se$_2$O & 2.86 & 1.23 & 1.34 & 2.55 &  1.40 \\
AB & 2.53 & 1.24 & 1.27 & 2.10 & 0.43 \\
AB' & 2.80 & 0.81 & 0.28 & 2.61 & 0.96 \\
\end{tabular}
\end{ruledtabular}
\end{table}

Next, we explore the optical selection rule in monolayer V$_2$S$_2$O. We found that the CBM valley at $X$ ($Y$) is almost 100\% contributed by the $d_{yz}$ ($d_{xz}$) orbital of V1 (V2), while the VBM valleys exhibit a hybrid nature, consisting of mainly $d_{xy}$ V1 (V2) atoms and minorly $p_y$ ($p_x$) from two S atoms. 
The little (local) group at $X$ ($Y$) is $D_{2h}$, in which the irreducible representations  (IRs) of VBM and CBM at $X$ ($Y$) are $B_{1g}$ ($B_{1g}$) and $B_{3g}$ ($B_{2g}$), respectively. Clearly, optical transitions between VBM and CBM at both $X$ and $Y$ are not allowed under this inversion symmetry, leading to the optical inactive (dark) excitons. On the other hand, optical transition are allowed along $\Gamma$-$X$ and $\Gamma$-$Y$, since the little group for these high-symmetric $\mathbf{k}$ points is $C_{2v}$.
Moreover, symmetry analysis shows that the IRs for the VBM and CBM along $\Gamma$-$X$ are $B_2$ and $A_2$, respectively. This means that the optical transitions are only allowed  when applied by the light fields with $y$ (with IR of $B_1$) directions, as $A_2 \bigotimes B_1 \bigotimes B_2 = A_1$. Vice versa, the optical transitions along $\Gamma$-$Y$ are only allowed with a $x-$polarized light. Therefore, there is a spin-momentum locking linear dichroism that spin-up (spin-down) electrons along $\Gamma$-$X$ ($\Gamma$-$Y$) are selectively excited, which provides a solid foundation for generating the optical active SML excitons in V$_2 X_2$O monolayer with linearly polarized light.
In Table~\ref{tab:exciton}, we collected the selected $\mathbf{k}$ points, active light polarization direction, orbital component of VBM and CBM, as well as the corresponding excitons that will be studied in the following.

\begin{table}[htb] 
\caption{\label{tab:exciton} In V$_2$S$_2$O monolayer, collected properties of $S$-th low-lying excitons, including coordinates $(k_x,k_y)$ of distributed $\mathbf{k}$ points ($k_z$ omitted), active light polarization directions $\sigma^{x/y}$, (mainly) orbital components of corresponding valence and conduction bands, respectively.} 
\begin{ruledtabular}
\begin{tabular}{rcccc}
  $S$ &  $\mathbf{k}$ & $\sigma^{x/y}$ & VB    &   CB  \\
\colrule
$D^{\uparrow}_1$   & $X$(0.5, 0.0)  & $y$    & $d_{xy}$@V1+$p_y$@S    &  $d_{yz}$@V1  \\
$D^{\downarrow}_1$ & $Y$(0.0, 0.5)  & $x$    & $d_{xy}$@V2+$p_x$@S    & $d_{xz}$@V2 \\
$E^{\uparrow}_1$   & ($\sim$0.2, 0.0) & $y$   & $p_z$@S+$p_x$@S    & $d_{yz}$@V1  \\
$E^{\downarrow}_1$ & \~(0.0, $\sim$0.2) & $x$  & $p_z$@S+$p_y$@S  & $d_{xz}$@V2  \\
\end{tabular}
\end{ruledtabular}
\end{table}

We then solved the BSE for the imaginary part of dielectric function and exciton energies for V$_2$S$_2$O monolayer based on the wavefunction by PBE+$U$. The calculated optical spectra along [110] direction are plotted in Figure~\ref{fig:2}(c), where the enlarged band gap of 3.28 eV by the scissors operator are indicated by the vertical solid line.
Comparing the optical spectra by BSE and independent-particle approximation (IP), we identify three main excitonic absorption peaks with significant optical responses below the shifted band gap and the absorption edge (around 3.8 eV) of the IP spectrum, indicating strong excitonic effects in the V$_2$S$_2$O monolayer. The highest absorption peak at about 3.1 eV corresponds to the $E_3$ exciton, which is largely contributed by the transitions off the highly polarized lines ($\Gamma$-$X$ and $\Gamma$-$Y$). This implies a much weaker spin selectivity in momentum-space of $E_3$, which is not the SML excitons we focus on in this work.

In Figure~\ref{fig:2}(d), we plotted the reciprocal-space distributions of optical transitions for several representative SML excitons, namely the bright excitons ($E^{\uparrow/\downarrow}_1$) and dark excitons ($D^{\uparrow/\downarrow}_1$) of spin-up/spin-down.
$E^{\uparrow}_1$, as well as the spin counterpart $E^{\downarrow}_1$, is the lowest energy (2.39 eV) bright exciton. 
The optical transition distributions in Figure~\ref{fig:2}(d) confirm a clear momentum-selectivity for $E^{\uparrow}_1$ and $E^{\downarrow}_1$, which couple to spin-up bands along $\Gamma$-$X$ and spin-down bands along $\Gamma$-$Y$, respectively.
In addition, the optical spectra by BSE with light polarization along $x$ and $y$ are plotted in the SM\cite{supp}. As expected, these two spectra are almost identical to each other, and the oscillation strength $\mu_S^2$ of SML excitons under light along either direction is highly spin-dependent. 
For instance, the oscillation strength of $E^{\uparrow}_1$ ($\mu_{E_1}^2(\uparrow)$) is almost 100 times larger than $\mu_{E_1}^2(\downarrow)$ when applied by the $y$-polarized light, and vice versa. Such an almost 100\% spin-polarization for $E^{\updownarrows}_1$ excitons, i.e. $\eta(E_1)=[\mu_{E_1}^2(\uparrow)-\mu_{E_1}^2(\downarrow)]/[\mu_{E_1}^2(\uparrow)+\mu_{E_1}^2(\downarrow)]$, shows a perfect SML linear dichroism as predicted above, 
which is distinct to the selective optical response to left- and right-handed circularly polarized light at $\bf K$ and $\bf K'$ in TMDCs\cite{Cao2012nc-TMDvalley}.

Moreover, compared to the A-excitons in TMDCs, our calculation shows that $E^{\uparrow}_1$ ($E^{\downarrow}_1$) has a much larger binding energy of 1.55 eV, implying a longer coherence time to recombine vertically.
However, one should notice that $E^{\uparrow}_1$ ($E^{\downarrow}_1$) is mainly contributed by the transitions from the saddle point of the VB near $\Gamma$ to the hillside of the CB valley along $\Gamma$-$X$ ($\Gamma$-$Y$).
Instead of being confined at the valley in reciprocal space as A-excitons in TMDCs, such a transition character will further result in a different carrier dynamics.
One could expect that the excited electrons of $E_1$ here are more likely to move downhill along $\Gamma$-$X$ ($\Gamma$-$Y$), whereas the holes would be tracked around the saddle points on the way from $\Gamma$ to the VBM without any external driving force.
Another effect of such a hillside-located electronic state is the ill-defined effective mass, leading to the invalidation of Eq.~\ref{tau_K} for the radiative lifetime at finite temperature. The estimated radiative lifetime of $E^{\uparrow}_1$ ($E^{\downarrow}_1$) at 0 K (using Eq.~\ref{tau_0}) is $\tau^{E_1}_0=2.49$ ps, which is longer than the A-exciton in TMDCs\cite{Palummo2015a,TMD-finite-ITL-exciton,Jiang2021-IXreview} mainly due to its lower oscillation strength.

On the other hand, the transitions from VBM to CBM are parity-forbidden as studied above.
Consequently, the corresponding excitons are found dark, i.e. spin-up polarized $D^{\uparrow}_1$ and spin-down polarized $D^{\downarrow}_1$, whose reciprocal-space distributions of optical transitions are mainly around $X$ and $Y$, respectively. Such an optical selection rule dominated by the spatial inversion symmetry, leading to dark SML excitons at valleys. Beside of the SML linear dichroism, it is  another significant difference between V$_2 X_2$O and TMDC monolayers. It is noteworthy that the the simple relationship $E_\mathrm{b}=\frac{3}{2N_\mathrm{g}}E_\mathrm{g}$ proposed in Ref.~\cite{Jiang-Duan2017-PRL} for estimating excitonic binding energies applies only to excitons formed between band edges (VBM and CBM). While this model fails for $E^{\uparrow}_1$ ($E^{\downarrow}_1$), (which involves non-edge transitions), it offers a good test for the dark $D^{\uparrow}_1$ ($D^{\downarrow}_1$) at 2.08 eV (see Fig. 2(c)). For V$_2$S$_2$O monolayer, $N_\mathrm{g}=4$ (four equivalent valleys), yielding an estimated $E_\mathrm{b}=\frac{3}{8}E_\mathrm{g}=1.23$ eV using the scissor-corrected gap $E_\mathrm{g}+\Delta_\mathrm{sc}=3.28$ eV. This agrees well with our first-principles calculated $E_\mathrm{b}=1.20$ eV for $D^{\uparrow}_1$ ($D^{\downarrow}_1$), supporting the validity of our $\Delta_\mathrm{sc}=2.0$ eV from the aspect of many-body Coulomb screening effect.

To further verify the validity of our first-principles results of the excitons, we numerically solve the radial equation of Eq.~\ref{MW-model} via finite-difference approach. Before solving this equation, one need to determine the effective mass ($\mu$) of the exciton, as well as the screening length ($r_0$). We listed the calculated $M^*$ and $r_0$ in Table~\ref{tab:wannier}. We considered two paths in the BZ, namely, $X-\Gamma$ and $X-M$. It shows that the effective masses of single-particles are highly anisotropic. We take the averaged effective masses of electrons and holes at $X$($Y$) to account for the reduced mass for excitons  ($M^*$) in Eq.~\ref{MW-model}. For the lowest $s$-state exciton, the lowest eigenvalue ($E^{\mathrm{W-M}}_{\mathrm{b}}$) is solved to be -1164 meV (mind the opposite sign). This is in good agreement with the one obtained via BSE for $D^{\uparrow}_1$ ($D^{\downarrow}_1$) (1.2 eV), since $D^{\uparrow}_1$($D^{\downarrow}_1$) is the lowest exciton taking place at  $X$($Y$) instead of $E_1$ as we studied above. Therefore, our cross-validation of the binding energy demonstrates the reliability of our calculations, while also implies the applicability of this wide-applied model to 2D altermagnetic semiconductors.

\begin{table*}[htb] 
\caption{\label{tab:wannier} The calculated effective masses for single-particles and excitons, as well as 2D polarizability $\alpha$ and the excitonic binding energies $|E^{\mathrm{W-M}}_{\mathrm{b}}|$ (in eV) via Eq.~\ref{MW-model}. The values of effective masses of excited electron ($m^*_e$, CBM) and hole ($m^*_h$, VBM) at $X$ valley along $\Gamma-X$ and $M-X$ directions. The averaged effective mass are taken by $m^*=\sqrt{m^*_{\Gamma-X}m^*_{M-X}}$. The reduced effective mass of exciton [$M^*=m^*_e m^*_h/(m^*_e + m^*_h)$] is calculated using the averaged effective masses of corresponding electron and hole. All the effective masses are in the unit of original electron mass ($m_e$).}
\begin{ruledtabular}
\begin{tabular}{lccccccccc} 
	 &  \multicolumn{3}{c}{$m^*_e$ (CBM)} &  \multicolumn{3}{c}{$m^*_h$ (VBM)} & $M^*$ & $r_0$ & $|E^{\mathrm{W-M}}_{\mathrm{b}}|$  \\
      &   $\Gamma-X$ & $M-X$ & average &   $\Gamma-X$ & $M-X$ & average &  & & \\
\colrule
  V$_2$S$_2$O &   1.73  &  0.84 & 1.26 & 1.91 & 0.54 & 0.97 & 0.55 & 33.6 & 1.16 \\
  V$_2$Se$_2$O &  1.73  &  0.85 & 1.21 & 1.94 & 0.37 & 0.85 & 0.50 & 41.1 &  0.96 \\     
  AB1 &      1.91  &  0.83 & 1.25 & 1.94 & 0.36 & 0.84 & 0.50 & 78.5 & 0.60 \\
\end{tabular}
\end{ruledtabular}
\end{table*}

\subsection{V$_2$S$_2$O/V$_2$Se$_2$O vdWH}

To achieve robust manipulation of carriers with longer coherence time and higher mobility, excitonic states are usually expected to be locate at the band minima (valleys). The dark excitons at $X$ ($Y$) valleys in the monolayers could turn bright once the spatial inversion symmetry is broken. For instance, one can build a so-called Janus structure by substituting S atom in one of the two sublayers with a Se atom, or stack different monolayers into a heterobilayer, i.e. the V$_2$S$_2$O/V$_2$Se$_2$O vdWH. In this work, we focus on the latter, for the sake of feasibility in experiment and promising advantages of forming interlayer excitons in vdWHs.

\begin{figure}[htb]
\includegraphics[width=8.5 cm]{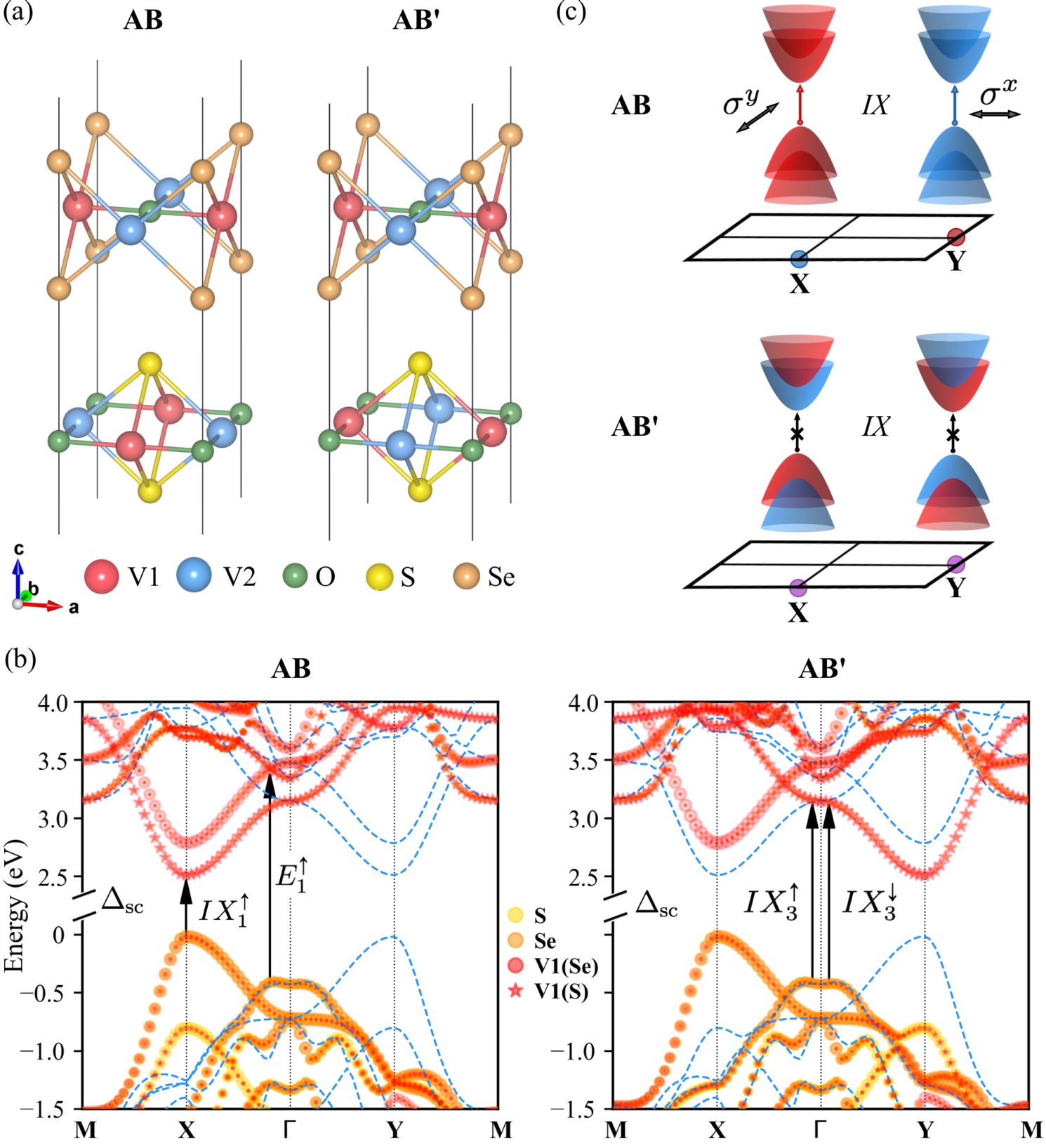}
\caption{\label{fig:hetero} The (a) geometry structures, (b) spin-resolved projected band structures by PBE+U with a scissors operator ($\Delta_\mathrm{sc}=2.0$ eV, indicated by the y-axis break) and (c) demonstration of optical selection rules of the interlayer excitations for AB and AB' stacked vdWHs. For a clear presentation, only the spin-up bands are projected onto the V1, S and Se atoms, while the spin-down bands are plotted with blue dashed lines for reference. The V1(S) and V1(Se) stands for the V1 (spin-up) atoms in the V$_2$S$_2$O and V$_2$Se$_2$O layers, respectively. The vertical arrows indicate the transitions of corresponding excitons.}
\end{figure}

We start with the most stable stacking order, i.e. AB-stacking in Figure~\ref{fig:hetero}(a).
The projected band structure of AB-stacking in Figure~\ref{fig:hetero}(b) exhibits an obvious Type-II band alignment, and the VBM and CBM are from V$_2$Se$_2$O and V$_2$S$_2$O monolayers, respectively.
In addition, the valleys of spin-up (spin-down) in the two monolayers are vertically aligned at $X$ ($Y$), meaning the VBM and CBM at each valley belong to the same spin-channel. The direct band gap, with the broken inversion symmetry, enables direct optical transition between VBM and CBM and therefore the formation of optically bright interlayer excitons at $X$ ($Y$) valleys.

\begin{figure}[htb]
\includegraphics[width=8.5 cm]{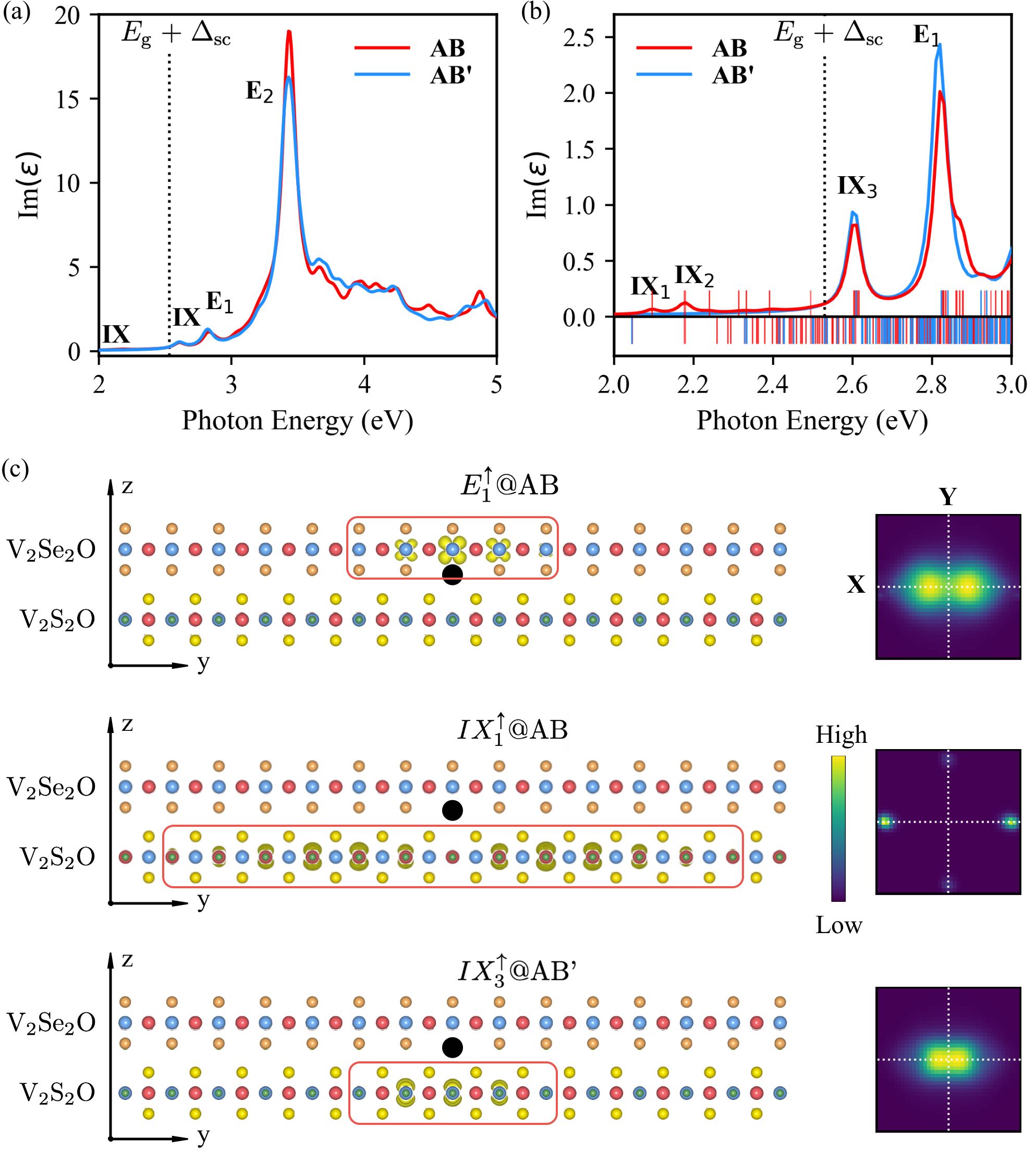}
\caption{\label{fig:hetero_exction} (a) The optical spectra Im($\varepsilon$) with the light polarization along [110] for AB and AB' vdWHs, color coded with red and blue, respectively. The single-particle band gaps of the whole vdWH after a rigid 2.0 eV scissors operator  ($E_\mathrm{g}+\Delta_\mathrm{sc}$) are indicated by the vertical dotted lines. (b) The detailed optical spectra to show the low-lying excitons. The vertical ticks in red(blue) indicate the excitonic energies for AB(AB')-stacking, with module square of dipole moment larger and smaller than $1\times10^{-2}$ are shown separately upon and below zero axis, respectively. 
(c) The real-space excitonic wavefunction modulus truncated at 2\% of their maximum values, and reciprocal-space distributions of optical transitions for the excitons $E1_1$@AB, $IX_1$@AB and $IX_3$@AB' in spin-up channel.}
\end{figure}

In Figure~\ref{fig:hetero_exction}(a) and (b), we plot the optical spectra of AB-stacking by BSE, as well as the eigenvalues for the excitons (see Figure~\ref{fig:hetero_exction}(b) with details). In the following, we denote the intralayer and interlayer excitons as $E^{\uparrow/\downarrow}_S$@$M$
and $IX^{\uparrow/\downarrow}_S$@$M$ in a particular system $M$ (e.g. AB for AB-stacking vdWH), respectively, where $S$ is the index of  excitons labeled in Figure~\ref{fig:hetero_exction}(a-b) and $\uparrow/\downarrow$ means spin direction of the exciton if needed.
The two dominant absorption peaks at 2.8 and 3.4 eV are assigned to the $E_1$@AB and $E_2$@AB, which are both intralayer excitons. As shown by the real-space excitonic wavefunction modulus ($|\psi(r_e, r_h)|^2$) in Figure~\ref{fig:hetero_exction}(c), both the electron and hole of $E_1$@AB distribute in the V$_2$Se$_2$O layer. The reciprocal-space distribution of transitions along $\Gamma$-$X$ for $E^{\uparrow}_1$@AB reveals its SML nature similar to the $E_1$ exciton in the V$_2$S$_2$O monolayer.
The $E_2$@AB is in fact the non-SML $E_3$ exciton in V$_2$S$_2$O monolayer (3.2 eV, see Figure~\ref{fig:2}).

The three peaks below $E_1$ are all interlayer excitons in AB-stacking, namely, $IX_1$, $IX_2$ and $IX_3$. We only discuss $IX_1$ here, since similar properties are found for $IX_2$ and $IX_3$ (see the SM\cite{supp}).
Our calculations show the dipole oscillator strength of $IX^{\uparrow}_1$@AB is mostly (over 99\%) contributed by the transitions between the VBM and CBM at $X$ valley, indicating intrinsically perfect spin-valley polarization during excitation. The $|\psi(r_e, r_h)|^2$ in Figure~\ref{fig:hetero_exction}(c) clearly demonstrates its interlayer distribution, where the hole is fixed in the V$_2$Se$_2$O layer and the electron resides in the V$_2$S$_2$O layer. This confirms that optically active SML interlayer excitons can indeed be generated at the $X$($Y$) valley in V$_2$S$_2$O/V$_2$Se$_2$O vdWH.
Furthermore, the BSE calculated binding energy of $IX_1$@AB is over 430 meV, which exceeds those in TMDC vdWHs by at least 25\%. Using the 2D Wannier-Mott model as above, we obtained a binding energy $|E^{\mathrm{W-M}}_{\mathrm{b}}|$ of 604 meV.
Note it is reasonable for the 2D Wannier-Mott model to overestimate the binding energy, since the vdWH is no longer a pure 2D dielectric environment.

In addition, the radiative lifetime at 0 K of $IX_1$ exciton is $\tau_0^{IX_1}=94.1$ ps. When taking averaged effective mass of the exciton , the $\tau^{IX_1}$(4K), $\tau^{IX_1}$(300K) in AB-stacking are estimated to be over 2.8 ns and 210 ns, respectively. Even when taking the effective masses along $M-X$, the $\tau^{IX_1}$(300K) is still over 107 ns,
demonstrating that significantly longer lifetimes of $IX_1$ could be expected in this altermagnetic vdWH against the conventional TMDC vdWHs at both low and room temperatures\cite{TMD-finite-ITL-exciton,Jiang2021-IXreview}.

Recently, the introduction of a rotational misalignment between the two layers in vdWHs offers a fascinating modulation of the interlayer excitons via the formation of local stacking-order dependent moir\'e potential\cite{Rivera2018-IX,Jiang2021-IXreview,Ciarrocchi2022}, establishing a new field termed {\it twistronics}. 
In this context, we consider another stacking order, i.e. AB'-stacking in Figure~\ref{fig:hetero}(a), which could be obtained by rotating one of the layers in AB-stacking by 90\textdegree~with a total energy increase of 10 meV per unit cell.
Apart from their spin configurations, AB- and AB'-stackings have the same geometry. Therefore, the two calculated band structures in Figure~\ref{fig:hetero}(b) are almost identical in energy.
However, the spin lattices (V atoms) of AB- and AB'-stackings exhibit antiferromagnetic and ferromagnetic alignments along the $c$-axis, respectively, leading to qualitatively different spin-dependent band structures.
In AB'-stacking, the 90\textdegree~rotation switches the $X$ and $Y$ in the original Brillouin zone of the operated monolayer (V$_2$S$_2$O here). As a result, the spin-valleys in the two monolayer are anti-aligned, meaning the VBM and CBM at each valley in AB'-stacking reside in opposite spin channels.

Consequently, such an anti-aligned SML configuration in AB'-stacking leads to the forbidden interlayer transitions at $X$($Y$) valleys due to the spin-flipping. As demonstrated by the optical selection rules in Figure~\ref{fig:hetero}(c), bright interlayer excitons ($IX$s) at $X$($Y$) could be selectively generated by linearly polarized light exclusively in AB-stacking, while those in AB'-stacking are dark in principle. As expected, although Figure~\ref{fig:hetero_exction}(a) shows roughly identical optical spectra of the two stackings, the main disparity between the spectra comes with the interlayer excitonic peaks in Figure~\ref{fig:hetero_exction}(b).
The spectrum of AB'-stacking exhibits no interlayer excitonic peaks below the intralayer $E_1$@AB', except an $IX_3$@AB'. Given the clear interlayer distributed $|\psi(r_e, r_h)|^2$, one may be surprised how this $IX_3$@AB' survives under the spin-flip restriction. In fact, $IX^{\uparrow}_3$@AB' are mostly generated by the optical transitions around $\Gamma$ point with a slight dispersion along $\Gamma$-$X$ (see Figure~\ref{fig:hetero} and \ref{fig:hetero_exction}), highlighting the distinct nature of $IX_3$@AB' against $IX_3$@AB at $X$($Y$) valleys. 
This $IX_3$@AB' cannot be regarded as SML excitons, since $IX^{\uparrow}_3$@AB' and its counterpart $IX^{\downarrow}_3$@AB' are too close to distinguish in the momentum-space.
Therefore, only $IX_1$, $IX_2$ and $IX_3$ in AB-stacking are the expected SML interlayer excitons with robust spin-valley coupling, indicating an effective tunability of the SML excitons upon the stacking order and a potential application in the twistronics.

\subsection{Discussions}

Apart from the large binding energy and long radiative lifetime, we now discuss the other advantages of SML excitons in altermagnetic V$_2 X_2$O systems over those in the TMDCs, which benefit directly from the large non-relativistic spin-splittings. In monolayer TMDCs, clear low-lying A and B excitonic peaks appear in the optical spectra\cite{Rivera2018-IX,Jiang2021-IXreview}, which could further evolve into four excitonic states, namely A-A' and B-B' after the full SOC effect is included\cite{Qiu-MoS2-prl2013,Shudong2024-prb,Yambo-SpinorExciton}.
The B exciton—with spins opposite to the lowest A exciton—stems from transitions within the same K valley between the spin-split valence band below the VBM and the first conduction band of matching spin. Due to the small magnitude of spin-splitting by SOC, the A (A') and B (B') excitonic peaks split in the spectra for about 150$\sim$450 meV (depending on the strength of SOC).
This proximity enables significant intra-valley exciton mixing\cite{TMD-exciton-expt0}, complicating the experimental identification and manipulation of specific spin-polarized excitonic states\cite{TMD-exciton-expt1,TMD-exciton-expt2,TMD-exciton-expt3}.
Conversively, the giant non-relativistic spin-splittings ($>1.2$ eV) around $X$ ($Y$) valleys prevent the formation of spin-opposite ``B-like" excitons at the same valley in altermagnetic V$_2 X_2$O systems, until the emergence of non-SML $E_3$ exciton (over 700 meV beyond $E_1$) at $\Gamma$. Consequently, low-lying SML excitonic peaks perfectly merge with their spin-opposite counterparts in the optical spectra (Figures 2 and 4) for both monolayer and vdWH.

Another advantage follows up with the possible spin dynamics and the improvement of spin-polarization of SML excitons in V$_2 X_2$O as proposed above. 
We take the lowest-energy interlayer exciton in AB-stacking ($IX_1$@AB) for example, in consideration of its robust spin-valley coupling and spatial separation induced long lifetime.
Mediated by exchange-interaction, SOC or phonons, there are several possible relaxation routines for $IX_1$ that lead to spin-valley depolarization, which could be simply classified into intra-valley and inter-valley spin-relaxations\cite{Yu2014,Glazov2015reviewExciton,Thygesen2016,Wang2018valleyRealx,Guo2019,JiangXiang2021}.
For the former, either or both carriers could vertically hop to the higher states of opposite spin, which cause the single or double (exchange) spin-flippings. According to Fermi's Golden Rule, the probabilities of transitions are, in principle, proportional to $1/\Delta^{\updownarrows}_{\mathrm{CB}}$ ($1/\Delta^{\updownarrows}_{\mathrm{VB}}$) or $1/(\Delta^{\updownarrows}_{\mathrm{CB}}\Delta^{\updownarrows}_{\mathrm{VB}})$ for the single or double spin-flippings, respectively.
The giant non-relativistic $\Delta^{\updownarrows}_{\mathrm{VB}}$ and  $\Delta^{\updownarrows}_{\mathrm{CB}}$) over 1.2 eV for AB-stacking (see Table~\ref{tab:energy}) significantly suppress the intra-valley relaxation, no matter which mechanism takes place.
The same principle applies for the inter-valley scattering processes with spin conservation, e.g., a spin-up excited electron further hopping from VBM@$X$ to a higher state of spin-up at $Y$ valley.
On the contrary, the SOC-induced splittings in TMDCs are down to tens of meV–the energy magnitude of room-temperature thermal fluctuation, which is a much smaller energy barrier to overcome.
The other inter-valley scattering with spin flipped usually requires a collaboration of strong SOC and electron-phonon coupling\cite{JiangXiang2021,Xu2021,Helmrich2021}, which could be further suppressed in V$_2 X_2$O systems thanks to their weak SOC.
Note that the inter-valley exchange interaction also play a crucial role in the depolarization\cite{1975SJETP,ShamLJ1993ExcitonSpin,Yu2014,Urbaszek2014,Guo2019}. However, a quantitative evaluation of this effect, as well as the phonon-SOC inter-valley spin-flipping, is beyond the scope of this work, which is definitely the direction of our future work via theoretical evaluation or computational simulations. 
Overall, with the depolarization suppressed in different ways more or less, we suggest that SML excitons with long-standing spin-polarization could widely exist in 2D altermagnetic semiconductors.

\section{conclusions}
In summary, we performed a theoretical investigation on the electronic band structures and optical properties of monolayer V$_2 X_2$O ($X=$ S, Se) and their van der Waals heterobilayer. Our first-principles calculations predict intrinsically SML excitons in 2D altermagnetic V$_2 X_2$O with large excitonic binding energies over 1400 meV in monolayers and 430 meV in vdWH, giant non-relativistic spin-splitting over 1.2 eV. 
This altermagnetic mechanism locks spin orientation to momentum without SOC, enabling room-temperature-stable spin-polarization.
Via analysis of the electronic structure and symmetry, optical selection rule in monolayers is explored that linearly polarized light selectively generates SML excitons.
Stacking-dependent symmetry breaking in vdWH further allows tunable bright/dark interlayer excitons.
These results establish 2D altermagnets as a novel platform for ultrafast opto-spintronics, offering deterministic spin manipulation and enhanced quantum coherence for advanced device applications.
\newline

{\it Note added}—Following the completion of this work, we became aware of studies on excitons in 2D AMs using group theory and first-principles calculations\cite{Cao-Zutic-2025AM-exciton-model2025,YangLi2025valleyselectiveExcitonic}.

\begin{acknowledgments}
This work was financially supported by the Ministry of Science and Technology of the People’s Republic of China (No.2022YFA1402901), the National Natural Science Foundation of China (NSFC No.T2125004, No.U24A2010,  No.12274228 and No.22303041), the NSF of Jiangsu Province (No.BK20230908), Fundamental Research Funds for the Central Universities (No.30922010102, No.30922010805 and No.30923010203), and funding (No.TSXK2022D002) as well as a startup grant from the NJUST. The authors acknowledge support from the Tianjing Supercomputer Centre, Shanghai Supercomputer Center, and High Performance Computing Platform of Nanjing University of Aeronautics and Astronautics.
\end{acknowledgments}


\bibliography{VXO-exciton}

\end{document}